\documentstyle[12pt,epsf]{article}

\newcommand{\plb}[3]{{{\it Phys.~Lett.}~{\bf B#1} (#3) #2}}
\newcommand{\npb}[3]{{{\it Nucl.~Phys.}~{\bf B#1} (#3) #2}}
\newcommand{\prd}[3]{{{\it Phys.~Rev.}~{\bf B#1} (#3) #2}}
\newcommand{\ptp}[3]{{{\it Prog.~Theor.~Phys.}~{\bf #1} (#3) #2}}

\newcommand{\leqsim}{\,\raisebox{-0.6ex}{$\buildrel < \over \sim$}\,}

\newcommand{\be}{\begin{equation}}
\newcommand{\ee}{\end{equation}}
\newcommand{\ba}{\begin{eqnarray}}
\newcommand{\ea}{\end{eqnarray}}

\def\etal{{\it et al}\,}
\def\hc{{\rm h.c.}}

\def\O{{\cal O}}

\def\gev{\,{\rm GeV}}

\begin{document}

\thispagestyle{empty}
\begin{flushright}
{\tt ULB-TH-96/15\\ hep-ph/9608251}
\end{flushright}
\vspace{5mm}
%\vspace{5cm}
\begin{center}
{\Large \bf Could the MSSM have no CP violation\\ in the CKM matrix? }\\
\vspace{15mm} {\large S.~A.~Abel and J.-M.~Fr\`ere}\\
%\footnote{feetnote}}\\
\vspace{1cm}
{\small\it Service de Physique Th\'eorique, 
Universit\'e Libre de Bruxelles\\
Boulevard du Triomphe, Bruxelles 1050, Belgium }
\end{center} 
\vspace{2cm}

\begin{abstract} 
%\begin{center}
\noindent We show that all CP violation in
the MSSM could originate in the supersymmetry breaking sector rather
than the CKM matrix, and discuss the important consequences for
$B$-physics. 
%\end{center}   
\end{abstract} 
\vspace*{1 cm}
%%{\bf PACS codes}: 11.30.Pb, 12.10.Dm, 12.60.Jv, 13.40.Em \\
%%{\bf Keywords}: Neutron electric dipole moment, minimal supersymmetric 
%%standard model \\

\newpage

\section{Introduction}

\noindent When discussing supersymmetric extensions to the Standard Model (SM),
most authors choose to incorporate the Kobayashi-Maskawa model of CP
violation~\cite{km}. In the Minimal Supersymmetric Standard Model
(MSSM), as in the SM, this can successfully explain the experimental
observations of CP violation (which admittedly are in rather short
supply).  However, there are many other possible sources of CP
violation in the MSSM, such as phases on trilinear $A$-couplings and
bilinear $B$-couplings. In fact writing the superpotential of the MSSM as
\begin{equation}    
\label{super_mssm}                                           
W=h_U Q_L H_2 U_R + h_D Q_L H_1 D_R
 + h_E L H_1 E_R + \mu H_1 \varepsilon H_2,
\end{equation}    
where generation indices are implied (and where the left-handed
superfields contain the antiparticles, with the VEVs of the Higgs
fields ($v_1$ and $v_2$) defined such that $m_u = h_U v_2$, $m_d = h_D
v_1$ and $m_e =h_E v_1$), CP violation can appear in any of the soft
supersymmetry breaking terms which consist of mass-squared scalar
terms, gaugino masses, and scalar couplings of the form, 
\begin{eqnarray}                                               
-\delta {\cal{L}}&=&m_{ij}z_i z_j^*  + \frac{1}{2}M_A
\lambda_A \lambda_A \nonumber\\ 
&& + A_U {\tilde q}^*_L h_2 {\tilde u}_R 
+ A_D {\tilde q}^*_L h_1 {\tilde d}_R
+ A_E {\tilde l}^* h_1 {\tilde e}_R + B \mu h_1 \varepsilon h_2  +\hc,
\end{eqnarray}    
where again, generation indices are suppressed, the $\lambda_A$ are the
gauginos, and the $z_i$ are generic scalar fields.
In the case that the couplings $A$, $M_A$ and $m_{ij}$ are all
degenerate at the GUT scale,
\ba
\label{degen}
A_{U_{ij}}&=&A h_{U_{ij}}\nonumber\\
A_{D_{ij}}&=&A h_{D_{ij}}\nonumber\\
A_{E_{ij}}&=&A h_{E_{ij}}\nonumber\\
m_{ij}   &=&\delta_{ij} m_0^2\nonumber\\
M_A      &=&m_{1/2},
\ea
there are four physical phases describing CP violation which were
ennumerated in refs.\cite{fg,dgh}.  Two of these are the usual $\theta
$ angle and CKM phase.  As pointed out in ref.\cite{fg}, only the
relative phases between $A$ and $B$ and $m_{1/2}$ are physically
significant since the phase on $m_{1/2}$ may be removed by a suitable
$R$-rotation. Thus the other two CP phases are those on $(A
m^*_{1/2})$ and $(B m^*_{1/2})$ (denoted $\phi_A$ and $\phi_B$
respectively).
   
Thus a scenario which is complimentary to the one usually considered,
is one in which CP violation arises {\em only} in the
soft-supersymmetry breaking terms, with the CKM matrix being entirely
real. In fact this possibility had earlier been considered in
ref.\cite{fg} for degenerate $A$, $M_A$ and scalar masses at the GUT
scale. Here it was found that the direct CP violation parameter,
$\varepsilon'$, was generally too large.  The subsequent work by Dugan
{\em et al} discouraged any further attempts in this direction, since
they placed quite severe limits on the values of $\phi_A$ and $\phi_B$
by using experimental bounds on the electric dipole moments (EDM) of
the neutron and electron. Typically one imposes
\be
\label{edmconst}
\phi_A{\mbox ,} \phi_B\leqsim \mbox{few}\times 10^{-3}.
\ee
Such small phases are unable (by themselves) to generate the CP
violation parameters ($\varepsilon$ and $\varepsilon'$) of the
$K-\overline{K}$ system.  The usual choice is to take these phases
instead to be exactly zero, in which case CP violation leaks into the
scalar couplings only through the running of the renormalisation group
equations. The resulting dipole moments are enhanced over those in the
SM, although probably not measurably so~\cite{inui,us}.

More recently, it has been demonstrated that, with the commonly 
adopted set of supersymmetry parameters, $\phi_A$ is far less
constrained than $\phi_B$~\cite{fo} (and we independently reproduce
these findings). This might give hope that the CP violation in the
$K$-system could arise purely from phases on the $A$-terms.  The
purpose of this paper therefore, is to reexamine whether the CP
violation could reside {\em only} in the soft-supersymmetry breaking
terms, and to what extent such a scenario would be `fine-tuned'.  In
the next section we show that, with {\em degenerate} $A$-terms at the GUT
scale, it is if fact not possible to generate sufficiently large
values of $\varepsilon$ because of cancellations that occur.

We then go on to consider more general forms for the soft
supersymmetry breaking.  Since in this context EDMs are generated from
flavour diagonal terms, and $\varepsilon $ from off diagonal terms,
one might expect that is possible to avoid bounds from EDMs (such as
those in eq.(\ref{edmconst})) whilst at the same time generating
reasonable values of $\varepsilon$, if rather than being degenerate,
the $A$ parameters have an off-diagonal `texture'.  In the light of
recent work on supersymmetry breaking in string theory, this is a
reasonably well motivated assumption. In fact, one of the properties
of the supersymmetry breaking in these theories is that there are only
large, non-trivial phases on the $A$ terms, precisely when one expects
there to be a high degree of non-degeneracy (that is when
supersymmetry breaking is dominated by the moduli rather than the
dilaton).  (In addition, since CP is an exact (discrete gauge)
symmetry of the string theory, its appearance in the Yukawa couplings
is not particularly easy to explain.)

We shall see that one can indeed explain the CP violation observed 
in the $K-\overline{K}$ system with a rather simple non-degenerate 
structure for the soft-supersymmetry breaking. We then go on to
discuss the expected pattern of CP violation in the $B-\overline{B}$ 
system in this picture.

First let us discuss the procedure we have used. This is based on the
very complete analyses of the `constrained' MSSM by Kane {\em et al}
\cite{kane} and Barger {\em et al} \cite{BBO94}.  As in ref.\cite{us},
we have used two loop RGE evaluation of gauge and Yukawa couplings and
have minimised the full one-loop Higgs potential to determine the
parameters $\mu$ and $B$, including contributions from matter and
gauge sectors, but retaining the full flavour dependence in the RGEs.
The process is as described in ref.\cite{us} except here of course we
must allow for more general choices of supersymmetry breaking
parameters at the GUT scale. This requires a few modifications:

The first is to the equations for the electric dipole moments, which
now receive significant {\em left-left} contributions from diagrams
involving one higgs vertex and one gauge vertex~\cite{KO92}.  Let us
define the diagonalisations of the mass matrices as follows,
\begin{eqnarray}
&\mbox{squarks : }& V_{\tilde{q}}^\dagger M^2_{\tilde{q}} V_{\tilde{q}} 
= m^2_{\tilde{q}}\nonumber\\
&\mbox{neutralinos : }& V_N^\dagger M_{N} V_N = m_{\chi^0}\nonumber\\
&\mbox{charginos : }& U_C^\dagger M_C V_C = m_{\chi^{\pm}},
\end{eqnarray}
where the squark mass-squared term is of the form
\be
(\tilde{q}_L^\dagger ,\tilde{q}_R^\dagger )
M^2_{\tilde{q}}
\left(
\begin{array}{c}
\tilde{q}_L \\
\tilde{q}_R \\
\end{array}
\right),
\ee
and
\ba
\label{masses}
M^2_{\tilde{u}} &=&
\left(
\begin{array}{cc}
m_U^2 + K \delta m^2_{\tilde{U}_{LL}}K^\dagger 
& v_2 K A_U +m_U \mu v_1/v_2)
\nonumber\\
 v_2 A^\dagger_U K^\dagger + m_U \mu v_1/v_2 
& m_D^2 + \delta m^2_{\tilde{U}_{RR}} 
\end{array}
\right) \nonumber\\
M^2_{\tilde{d}} &=&
\left(
\begin{array}{cc}
m_D^2 + \delta m^2_{\tilde{D}_{LL}} 
& v_1 A_D +m_D \mu v_2/v_1
\nonumber\\
 v_1 A^\dagger_D + m_D \mu v_2/v_1 
& m_D^2 + \delta m^2_{\tilde{D}_{RR}} 
\end{array}\right),
\ea
and where the $\delta m^2 $ contain the renormalised squark
mass-squared terms and also generation independent contributions from
the $D$-terms.  We are using the down-quark diagonal basis, and $K$ is
the CKM matrix ($m_U=\mbox{diag}( m_u,m_c,m_t ) = K h_U v_2 $ ).  We
find the following chargino contributions to the quark electric dipole
moments\footnote{this corrects eq.(23) of ref.\cite{us} in which the
quark charges were omitted};
\begin{eqnarray}
d_d&=&-\frac{1}{3}\frac{e}{32 \pi^2} 
\sum_i^2 
\frac{(V_C)_{2i}(U_C)^*_{\alpha i}}{m_{\chi_i^\pm}}
\mbox{Im}
\left(
\delta_{\alpha 2} h^\dagger_D K^\dagger \left[ V_{\tilde{u}} 
F_d\left(\frac{m_{\tilde{u}}^2}{m_{\chi_i^\pm}^2}\right)
 V_{\tilde{u}}^\dagger \right]_{LR} h^\dagger_U \right.\nonumber\\
&& \left. \hspace{6cm}-\delta_{\alpha 1} 
h^\dagger_D K^\dagger \left[ V_{\tilde{u}} 
F_d\left(\frac{m_{\tilde{u}}^2}{m_{\chi_i^\pm}^2}\right)
 V_{\tilde{u}}^\dagger \right]_{LL} K g_2 \right)_{11}\nonumber\\
d_u&=&\frac{2}{3}\frac{ e}{32 \pi^2}  
\sum_i^2 
\frac{(V_C)_{\alpha i}(U_C)^*_{2i}}{m_{\chi_i^\pm}}
\mbox{Im}
\left(
\delta_{\alpha 2} h_U^\dagger \left[ V_{\tilde{d}} 
F_u\left(\frac{m_{\tilde{d}}^2}{m_{\chi_i^\pm}^2}\right)
 V_{\tilde{d}}^\dagger \right]_{LR} h_D^\dagger K^\dagger 
\right.\nonumber\\
&& \left. \hspace{6cm} -\delta_{\alpha 1} 
h_U^\dagger \left[ V_{\tilde{d}} 
F_u\left(\frac{m_{\tilde{d}}^2}{m_{\chi_i^\pm}^2}\right)
 V_{\tilde{d}}^\dagger \right]_{LL} K^\dagger g_2
\right)_{11},
\end{eqnarray}                        
where we have defined the functions, 
\begin{eqnarray}
F_d&=&\frac{1}{(1-x)^3}
\left[ 5-12x+7x^2+2x(2-3x)\log x\right] \nonumber\\
F_u&=&\frac{1}{(1-x)^3}
\left[ 2-6x+4x^2+x(1-3x)\log x\right].
\end{eqnarray}                        
For the case we are considering, the CKM matrix will of course be
real.  For the gluino contributions we find, 
\begin{eqnarray}
d_d&=&-\frac{e \alpha_s}{9 \pi m_{\tilde{g}}} 
\mbox{Im}\left(
\left[ V_{\tilde{d}}
G\left(\frac{m_{\tilde{d}}^2}{m_{\tilde{g}}^2}\right)
 V_{\tilde{d}}^\dagger \right]_{LR}\right)_{11}\nonumber\\
d_u&=&\frac{2 e \alpha_s}{9 \pi m_{\tilde{g}}} 
\mbox{Im}\left(
\left[ V_{\tilde{u}}
G\left(\frac{m_{\tilde{u}}^2}{m_{\tilde{g}}^2}\right)
 V_{\tilde{u}}^\dagger \right]_{LR}\right)_{11},
\end{eqnarray}                        
where we have defined the function 
\be
G=\frac{1}{(1-x)^3}
\left[ 1-x^2+2x\log x\right].
\ee
The neutralino contributions, which were also included, were found 
always to be small. 

The second modification is to the conditions which indicate whether
the minimum obtained is global, or whether there are other minima
which may have broken colour or charge (CCB), or directions in which
the potential is unbounded from below (UFB).  Necessary conditions
were deduced in refs.\cite{color}, and have been exhaustively
generalised in refs.\cite{casas,cd}. Since here we are considering the
possibility of large non-degeneracy in the $A$-terms, it is especially
important to use the flavour violating conditions of ref.\cite{cd}
which take a particularly simple form. The CCB conditions are,
\ba
|A_{U_{ij}}|^2 &\leq & |h_{U_{kk}}|^2 \left( m_{uL_i}^2 +
m_{uR_j}^2+m_2^2+\mu^2 \right) \nonumber\\
|A_{D_{ij}}|^2 &\leq & |h_{D_{kk}}|^2 \left( m_{dL_i}^2 +
m_{dR_j}^2+m_1^2+\mu^2 \right) \nonumber\\
|A_{E_{ij}}|^2 &\leq & |h_{E_{kk}}|^2 \left( m_{eL_i}^2 +
m_{eR_j}^2+m_1^2+\mu^2 \right), 
\ea
where $i\neq j$, $k=\mbox{Max}(i,j)$ and $m_1^2$ and $m_2^2$ are 
the scalar mass-squared terms for the higgs, and the UFB conditions
are,  
\ba
|A_{U_{ij}}|^2 &\leq & |h_{U_{kk}}|^2 \left( m_{uL_i}^2 +
m_{uR_j}^2+m_{eL_p}^2+m_{eR_q}^2 \right) \nonumber\\
|A_{D_{ij}}|^2 &\leq & |h_{D_{kk}}|^2 \left( m_{dL_i}^2 +
m_{dR_j}^2+m_{\nu_m}^2 \right) \nonumber\\
|A_{E_{ij}}|^2 &\leq & |h_{E_{kk}}|^2 \left( m_{eL_i}^2 +
m_{eR_j}^2+m_{\nu_m}^2 \right), 
\ea
where $p\neq q$ and $m\neq i\neq j $. For the diagonal terms we used 
the more complete expressions given in ref.\cite{casas}.

The $\varepsilon $ parameter was calculated using the expressions for
the MSSM of refs.\cite{goto,gabbiani}. Since the SM contributions are
insignificant here (see below), the main contributions are from
chargino and gluino box diagrams. To demonstrate our nomenclature, we
shall present the full chargino terms for left-handed external quarks
here.  The contributions to the mixing matrix elements are as follows;
\ba
\label{goto}
M_{12} (K) &=& \frac{B_K \eta_K f_K^2 M_K}{384\pi^2}\left[
A_{SM}+A_{H^\pm}+A_{\chi^\pm}+A_{\tilde{g}} \right]\nonumber\\
A_{\chi^\pm} &=& \sum^2_{\alpha\beta}\sum^6_{ij}
\frac{g_2^4}{m^2_{\chi_\alpha}}
\left[ g_2 (V^\dagger_{\tilde{u}L} K)^i_2 (V_{C})^1_\alpha-
(V_{\tilde{u}R}^\dagger h_U^\dagger )^i_2 (V_{C})^2_\alpha \right]
\nonumber\\
&& \times 
\left[ g_2 (V^\dagger_{\tilde{u}L} K)^j_2 (V_{C})^1_\beta-
(V_{\tilde{u}R}^\dagger h_U^\dagger )^j_2 (V_{C})^2_\beta \right]
\nonumber\\
&& \times 
\left[ g_2 (K^\dagger V_{\tilde{u}L})^1_i (V^\dagger_{C})^1_\beta-
(h_U V_{\tilde{u}R} )^1_i (V^\dagger_{C})^2_\beta \right]
\nonumber\\
&& \times 
\left[ g_2 (K^\dagger V_{\tilde{u}L})^1_j (V^\dagger_{C})^1_\alpha-
(h_U V_{\tilde{u}R} )^1_j (V^\dagger_{C})^2_\alpha \right]
\nonumber\\
&& \times 
\hat{F}(m_i^2/m_{\chi^\pm}^{\alpha 2},m_j^2/m_{\chi^\pm}^{\alpha 2},
m_{\chi^\pm}^{\beta 2}/m_{\chi^\pm}^{\alpha 2} )
\ea
where we have defined the $6\times 3 $ matrices
$(V_{\tilde{q}L})_i^a=(V_{\tilde{q}})_i^a$, and
$(V_{\tilde{q}R})_i^a=(V_{\tilde{q}})_i^{a+3}$, and where $\hat{F}$
represents combinations of Inami-Lim functions\cite{goto}. (The terms
with right-handed quarks are expected to be insignificant for the
charginos since they are suppressed by Yukawa couplings.) For the
gluino contribution $A_{\tilde{g}}$ we used the approximations of
ref.\cite{gabbiani} which include all chiralities of external quarks.

The mass-insertion approximation was also used for the $\varepsilon '$
parameter (see ref.\cite{gabbiani} and references therein). In view of
other uncertainties this was sufficient for the present analysis.
Other possible FCNC effects were also checked using the expressions of
ref.\cite{gabbiani}, except for $b\rightarrow s\gamma $, for which the
full expressions of ref.\cite{BBM} were used.

\section{The Degenerate MSSM}

\noindent Before considering more general soft supersymmetry breaking, we shall
first discuss the effect of having degenerate boundary conditions as
in eq.(\ref{degen}), but following ref.\cite{fo} allow the phases
$\phi_A$ and $\phi_B$ to be non-zero. In this case it is not possible
to generate the experimentally observed CP violation if there is no CP
violation in the CKM matrix.

The reason why becomes apparent when one considers the leading
supersymmetric box-diagrams.  Consider for example the potentially
significant contribution to $\varepsilon $ from the chargino/up-squark
box with external left-handed quarks.  This diagram may be
approximated by the box-diagram with a single mass-insertion,
$M^2_{\tilde{u}LR} $, on the squark lines, and top-quark Yukawa
couplings, $h_U$ on two vertices. The contribution is of the form
\be
\varepsilon \propto \mbox{Im} \left( (h_U M^2_{\tilde{u}RL} K)_{12}^2 +
(K^\dagger M^{2}_{\tilde{u}LR}h^\dagger_U)_{12}^2 \right).
\ee
This corresponds to the cross-term in $A_{\chi^\pm}$ of
eqn.(\ref{goto}) when the Inami-Lim functions are expanded and the
leading linear terms taken. Since we are assuming no CP violation in the 
CKM matrix then $K^\dagger=K^T$ and $h_U^\dagger=h_U^T$. It is
convenient to define the matrices ${\bf A}_U$ such that $A_U= h_U
.{\bf A}_U $. The degenerate boundary condition corresponds to ${\bf
A}_{Uij}= A\delta_{ij} $, and 
\ba
\varepsilon \propto  \left( (h_U {\bf A}^\dagger_U
h_U^T)_{12}^2 -(h_U {\bf A}^*_U h_U^T)_{12}^2 
+ (h_U {\bf A}_U h^T_U)_{12}^2 -
(h_U {\bf A}^T_U h_U^T)_{12}^2\right).
\ea
In the event that the ${\bf A}_U$ matrix is symmetric, this contribution
completely vanishes. Inspection of the renormalisation group
equations (see for example ref.\cite{BBM}) shows that for degenerate 
boundary conditions this is the case to leading order. The matrix
${\bf A}_U$ is in fact found to be symmetric to typically one part 
in $10^{4}$ at the weak scale. 

This greatly suppresses any contribution to $\varepsilon $ from
chargino box diagrams, and similar arguments apply to the other box
diagrams too. In order to demonstrate this we shall consider a
`typical' point in parameter space where $A=500 \gev$, $m_0=300\gev$,
$m_{1/2}=100\gev$, $\tan \beta= 5$ and $\mu +ve$. Minimising the
effective potential gave the values $B=-116\gev$, and $\mu =
187\gev$\footnote{These results were verified using a different
minimisation routine to within $\pm 10\gev $ by P.~L.~White.}.  The
dependence of the EDM of the neutron on $\phi_A$ and $\phi_B$ is shown
in fig.(1). The contour $1.1 \times 10^{-25}$ clearly
agrees with the results in ref.\cite{fo}. In the region which is shown
in the plot, the value of $\varepsilon $ was never found to exceed
$2\times 10^{-11}$. (Note that this suppression occurs because the CKM
matrix is real; if one allows the usual CP phase into the CKM matrix, the
supersymmetric contribution to $\varepsilon $ is $\O (10^{-4} )$.)

\section{More General Parameters}

\noindent Before presenting some more general patterns of soft supersymmetry
breaking, let us say a little about how non-degenerate supersymmetry
breaking can arise in string theory. Recent progress in this area has
shown that the $A$-term for a coupling, $h_{ijk}$, between three
superfields, $ijk$, may at tree-level be written schematically in the
form \cite{softsusy,christoph,bailin}
\be
A_{ijk} \sim -m_{1/2} (1 + e^{i (\gamma_T)}\cot \theta 
\hspace{1mm} F_{ijk}) h_{ijk},
\ee
where the angle $\theta $ describes the goldstino direction, and where
the VEV of the dilaton is assumed to be real. When $\theta = \pi/2$
the supersymmetry breaking is along the dilaton direction, and when
$\theta=0 $ it is in the direction of moduli describing the size and
shape of the compactification.  The phase on the second term is the
putative source of CP violation and represents CP violation in the
VEVs of the moduli. Such spontaneous breaking of CP by moduli has been
discussed for orbifolds in ref.\cite{bailin}.The function $F_{ijk} $
is a function of the moduli VEVs and vanishes in a number of
interesting cases outlined in ref.\cite{christoph}. The first case is
obviously when supersymmetry breaking is dominated by the dilaton and
$\cot \theta=0$. However it is also clear that in this case $\phi_A =0
$ and the soft supersymmetry breaking cannot be the source of CP
violation.  The moduli dependent term also vanishes for renormalizable
couplings in which all the fields come from untwisted sectors, or have
weight -1 under certain duality transformations (for instance all
renormalizable couplings in the $Z_2\times Z_2$ orbifold satisfy this
criterion).  Thus one can identify a number of possibilities for
generating an off-diagonal structure in the $A$-terms, all of which
require supersymmetry breaking to be dominated by the moduli with 
$\cot\theta \gg 1$;

\begin{itemize}

\item The off diagonal Yukawa couplings come from non-renormalizable 
terms whereas the diagonal ones are renormalizable. 

\item The non-degeneracy is generated by 1-loop corrections, with the 
$A$-terms being zero at tree-level. This possibility has been
discussed recently in the context of FCNCs in ref.\cite{nir}.

\item The non-degeneracy is generated for couplings involving fields
with weights other than -1 (for example in the third generation only). 

\end{itemize}

These possibilities, together with the recent observation that the
pure dilaton breaking scenario breaks charge and colour~\cite{casas2},
make the assumption of non-degeneracy a reasonable one.

It is beyond the scope of this paper to discuss soft supersymmetry
breaking in string theory in any great depth, and we shall instead
select a number of `textures' to analyse. Here our aim will be merely
to demonstrate the possibility that CP violation comes only from the
soft supersymmetry breaking. As we shall see in the next section, the
experimental signatures of this scenario are quite striking, so that
for the moment they are of more immediate interest.

In order to anticipate the effect of various patterns of soft
supersymmetry breaking, it is useful to think in terms of the leading
mass insertion approximations. It is customary to consider the
parameters
\be
\delta^q_{ij} = \frac{M_{\tilde{q}_{ij}}^2 }{\tilde{m}^2}
\ee
where $\tilde{m}$ is an `average' sfermion mass. From the limits
derived in ref.\cite{gabbiani} it is clear which are the important
elements corresponding to each process provided that the gluino
diagrams are the dominant contribution.  The EDMs of the neutron and
electron impose quite severe limits on the imaginary diagonal
components in the left-right blocks, $(\delta^d_{11})_{LR}$,
$(\delta^u_{11})_{LR}$ and $(\delta^e_{11})_{LR}$.  The flavour
changing neutral currents on the other hand impose bounds on the
off-diagonal components; $b\rightarrow s\gamma$ constrains
$(\delta^d_{23})_{LR}$ and $(\delta^d_{23})_{LL}$ (weakly) and $\Delta
m_K $ depends on $(\delta^d_{1i})_{LL}$, $(\delta^d_{1i})_{LR}$, and
$(\delta^d_{1i})_{RR}$ where $i\neq 1$. Large values of these should
be avoided, although $\Delta m_K $ must inevitably be affected.  The
parameters $\varepsilon $ and $\varepsilon' $ depend most strongly on
$(\delta^d_{12})_{LL}$ and $(\delta^d_{13})_{LR}$.  There are however
relatively few constraints on $(\delta^u_{i\neq j})$ and in addition
$h_D$ is almost diagonal at the GUT scale. If we maintain the
assumption that the $A$-terms include factors of the Yukawa couplings,
this suggests that in the basis we are using, off-diagonal terms in
$M^2_{\tilde{d}}$ should be generated radiatively from terms in $A_U$.
 
We shall therefore consider the following `textures' for the 
$A$-matrices and the squark masses at the GUT scale, 
\ba
A_{U_{ij}}&=& A h_{U_{ij}}
+
\left(
\begin{array}{ccc}
0 & \delta A_{12} h_{U_{12}} & \delta A_{13} h_{U_{13}}\nonumber\\
\delta A_{21} h_{U_{21}} & 0 & \delta A_{23} h_{U_{23}}\nonumber\\
\delta A_{31}h_{U_{31}} & \delta A_{32}h_{U_{32}} & 0 
\end{array}
\right)\nonumber\\
A_{D_{ij}}&=&A h_{D_{ij}}\nonumber\\
A_{E_{ij}}&=&A h_{E_{ij}}\nonumber\\
m_{ij}    &=&\delta_{ij} m_0^2 + \delta m^2 \nonumber\\
M_A       &=&m_{1/2}\nonumber\\
\phi_B    &=&0.
\ea
The parameter $\delta m^2 $ represents off diagonal terms which may
also be generated in the mass-squared matrices.  From now on we shall
impose $\phi_B=0$ to avoid large EDMs, assuming that an explanation
for this lies in the mechanism which generates the 
$\mu$-term~\cite{softsusy,bailin}.

For simplicity we shall introduce the real parameters $\delta A$, and 
$\phi_{\delta A}$, and consider the following three symmetric structures;
\be
\label{texturesa}
A_{U_{ij}}= A h_{U_{ij}}
+
\delta A e^{i \phi_{\delta A}}\left(
\begin{array}{ccc}
0 & h_{U_{12}} & 0 \nonumber\\
 h_{U_{21}}& 0 & 0 \nonumber\\
0 & 0 & 0 
\end{array}
\right)
\ee
\be
\label{texturesb}
A_{U_{ij}}= A h_{U_{ij}}
+
\delta Ae^{i \phi_{\delta A}}\left(
\begin{array}{ccc}
0 & 0 & h_{U_{13}} \nonumber\\
0 & 0 & 0 \nonumber\\
h_{U_{31}} & 0 & 0 
\end{array}
\right)
\ee
\be
\label{texturesc}
A_{U_{ij}}= A h_{U_{ij}}
+
\delta Ae^{i \phi_{\delta A}}\left(
\begin{array}{ccc}
0 & 0 & 0 \nonumber\\
0 & 0 &  h_{U_{23}}\nonumber\\
0 &  h_{U_{32}} & 0 
\end{array}
\right).
\ee
For each of these possibilities there is a 7-dimensional parameter
space consisting of ($A$, $m_0$, $m_{1/2}$, $\tan\beta $, $\delta
m^2$, $\delta A$, $\phi_{\delta A}$) in addition to the sign of $\mu
$. The results are shown in figs.(2)-(4) for
$\phi_{\delta A}=\pi/4$\footnote{This is the maximal case. Smaller
values of $\phi_{\delta A}$ may be compensated by larger values of
$\delta A$.  For {\em very} small values such as those considered in
ref.\cite{bailin}, this may be considered to be a fine-tuning in the
sense that for the example of string derived soft terms, one requires
the goldstino to have almost no dilaton component.}.  The vertical
bounds in these figures are from CCB and UFB constraints, and the
horizontal bounds are from $\Delta M_K $ constraints.
 
As one might expect, the first texture is not very efficient at
generating $\varepsilon $ (since the relevant contribution in the RGEs
is Cabibbo suppressed), however there is a large region in each of the
remaining two cases which can successfully explain the observed value
of $\varepsilon $ whilst avoiding all other experimental constraints.
In addition the value of $\varepsilon'$ was in each case found to be
very small;
\be
\left|\frac{\varepsilon'}{\varepsilon}\right|\leqsim 10^{-6}
\ee
along the $\varepsilon = 2.3\times 10^{-3}$ contour.  In this sense
the experimental signatures are expected to be `superweak' with no
observable direct CP violation. The picture of CP violation here is
therefore more consistent with the results on $\varepsilon'$ from E731
than those from NA31 (see for example ref.\cite{roy} and references
therein).

For $B$-physics the picture is rather unusual. In $B$-physics, because
of the similar decay times of the two eigenstates, one cannot
disentangle the direct and indirect CP violation using just one
process. Instead one compares CP violation for different processes
using the parameters~\cite{big}
\be
\Phi_{CPV}(f)=\mbox{arg}\left( \frac{q}{p}\overline{\rho}(f) \right), 
\ee
where $q/p$ is assiciated with the mixing between $B^o-\overline{B}^o$
given by 
\be
\frac{q}{p} = \sqrt{ \frac{M_{12}^* - i \Gamma_{12}^*}
                          {M_{12} -   i \Gamma_{12}  }} ,
\ee
where $|q/p|\approx 1$. The parameter $\overline{\rho}(f)$ is related
to the direct CP violation in the decay $B\rightarrow f$.  Neither
$q/p$ nor $\overline{\rho}$ is phase-reparameterisation invariant, and
thus cannot be independently observed.  

In the SM, the $\overline{\rho}(f)$ receive contributions from
tree-level $W$-exchange diagrams, and different phases from the KM
matrix appear according to the channel considered, leading to a
determination of the angles in the unitarity triangle~\cite{big}.  The
pattern of CP violation here is in sharp contrast, since contributions
to direct CP violation arise only through penguin diagrams which in
addition to being one-loop, are suppressed by factors of Yukawa
couplings.  The {\em relative} phases of the various $\overline{\rho }
(f)$ are thus small with respect to the SM, and the
picture of CP violation is close to that of the `superweak' models in
the tree-level approximation. (One loop penguin diagrams may be
significant for processes which are Cabibbo suppressed at tree-level.)
There is therefore a basis (i.e. the one which we are using) in which all
the $\overline{\rho}(f)$ are approximately real for every process and
hence all the $\Phi_{CPV}$ are given by
\be
\Phi_{CPV}\approx -\mbox{arg}\left( M_{12}\right) . 
\ee
Moreover we find that, for the three examples studied here,
this phase is insignificant, in accord with 
previous analyses of the $B$-system in the constrained MSSM~\cite{goto}. 
Thus one concludes that {\em for the $B$-system there is little
detectable CP violation}. (Some higher loop contributions 
such as finite contributions to the Yukawa couplings, may be 
detectable for some Cabibbo suppressed processes.)

\section{Conclusion}

\noindent We have shown that the experimentally 
observed CP violation could be generated in the soft-supersymmetry
breaking sector of the MSSM rather than the Yukawa couplings. It is
possible to avoid constraints from EDMs and FCNCs by choosing an
off-diagonal texture for the trilinear couplings. The experimental
signatures of this type of CP violation are markedly different from
those in the SM or the `constrained' MSSM. Generally the CP violation
is expected to be of the `superweak' variety, arising only through
mixing, with little direct CP violation. For the $B$-system the
relatively small contribution to mixing means that there will be 
no detectable CP violation at all (modulo possible one-loop effects). 

This picture seems an attractive prospect for a number of reasons.
For example, if this scheme is correct, then in conjunction with other
FCNC processes, one has access to rather direct information about
physics occuring at the Planck scale, specifically the nature of the
supersymmetry breaking fields and their VEVs.

Another promising aspect is that of baryogenesis. In order to generate
a sufficient baryon number in the SM and even the MSSM, one generally
requires additional CP violation beyond that in the CKM matrix.  Here
however the CP violation responsible for the value of $\varepsilon $
could easily be sufficient to generate the observed baryon number
since it is a `hard' violation; CP violation in the SM for example is
typically suppressed at $T\gg m_t $ by a factor $\O (
m_{quark}^{12}/T^{12} )$ whereas here the suppression need only be $\O
(\tilde{m}^2/T^2 )$. This will be the subject of future work.\\

\section*{Acknowledgments}
\noindent It is a pleasure to thank P.~L.~White and I.~B.~Whittingham
for discussions, for help with computing, and for providing
cross-checks for some of the results presented here. This work was
supported in part by the European Flavourdynamics Network
(ref. chrx-ct93-0132).

\newpage

\begin{figure}
\label{edmfig}
\vspace*{-1in}
\hspace*{-1.0in}
\epsfysize=7.21in
\epsfxsize=7.21in
\epsffile{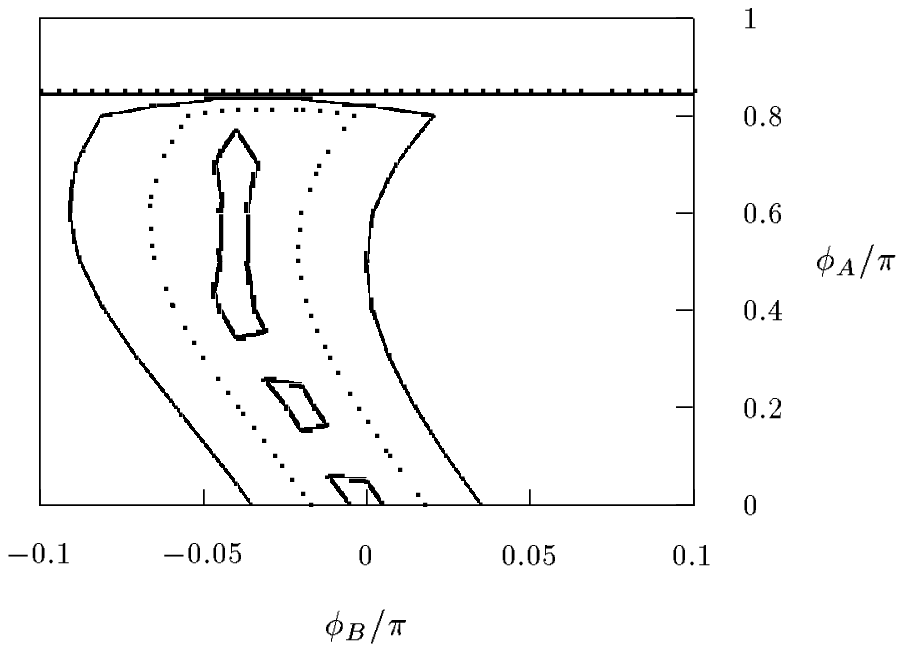}
\vspace{-3in}
\caption{\it The EDM of the neutron for the degenerate case with $A=500
\gev$, $m_0=300 \gev$, $m_{1/2}=100 \gev$ and $\tan \beta =5$ with 
$\mu =+ ve$. The contours are $1.1 \times 10^{-25}$ (thick-solid),
$5\times 10^{-25}$ (dotted) and $10^{-24} e$cm (solid). The jagged
line dilineates the region above which one cannot find a minimum.
The other constraints were not imposed for this diagram.}
\end{figure}
\begin{figure}
\label{epsfig1}
\vspace*{-1in}
\hspace*{-1.0in}
\epsfysize=7.21in
\epsfxsize=7.21in
\epsffile{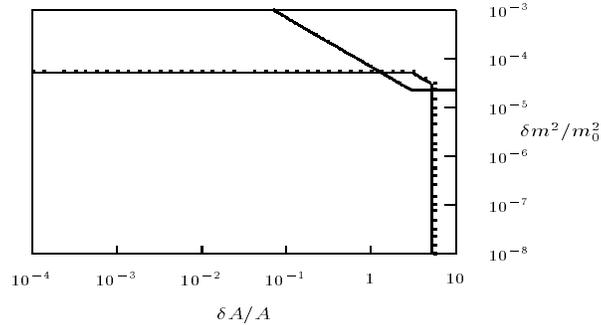}
\vspace{-3in}
\caption{\it The allowed ($\delta m^2 $, $\delta A $) parameter space, for
eq.(19). The allowed region is below the jagged line. The solid line is
the contour $\varepsilon = 2.3 \times 10^{-3}$ }
\end{figure}
\begin{figure}
\label{epsfig2}
\vspace*{-1in}
\hspace*{-1.0in}
\epsfysize=7.21in
\epsfxsize=7.21in
\epsffile{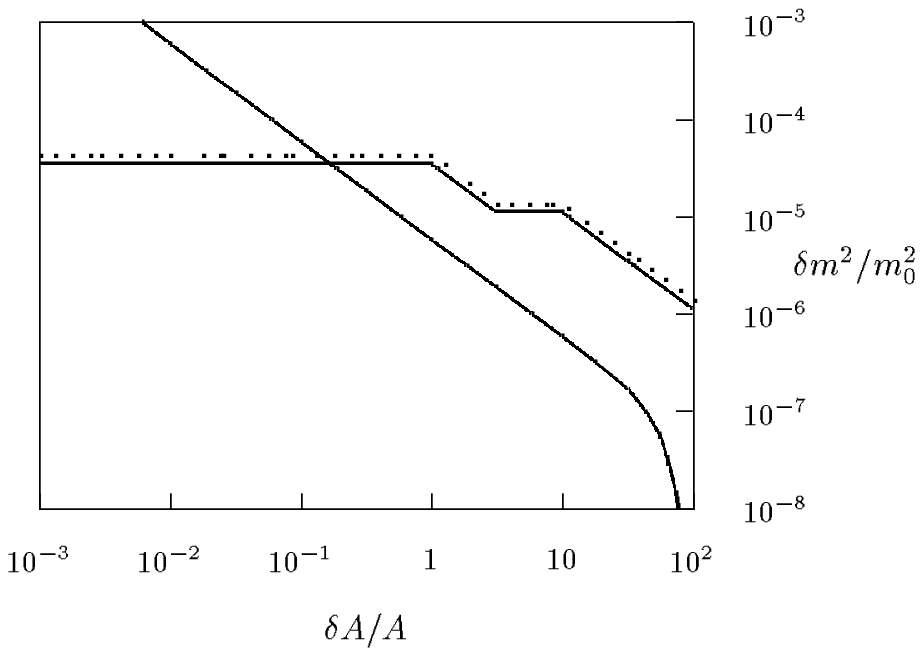}
\vspace{-3in}
\caption{\it The allowed ($\delta m^2 $, $\delta A $) parameter space, for eq.(20).}
\end{figure}
\begin{figure}
\label{epsfig3}
\vspace*{-1in}
\hspace*{-1.0in}
\epsfysize=7.21in
\epsfxsize=7.21in
\epsffile{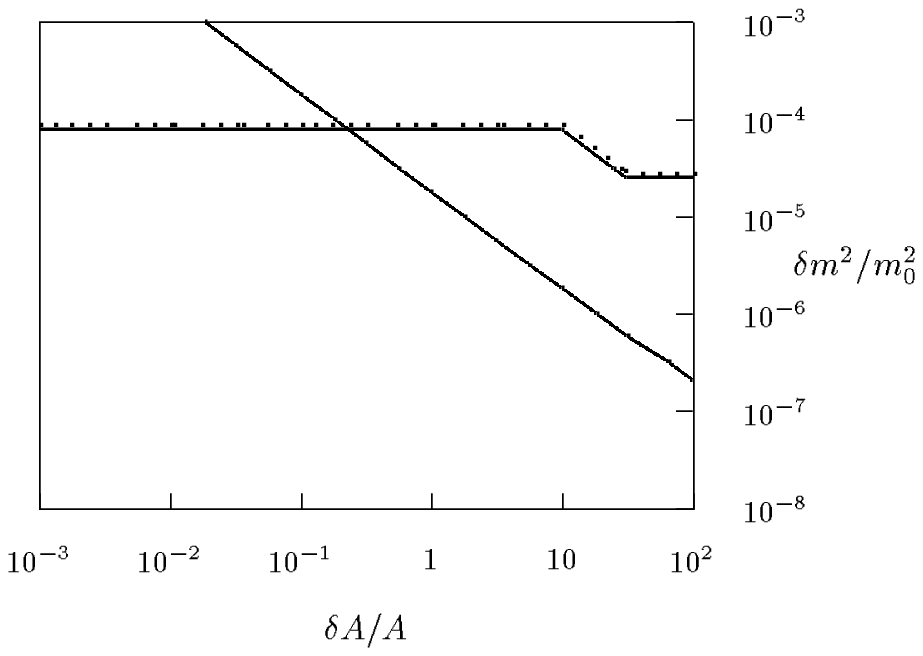}
\vspace{-3in}
\caption{\it The allowed ($\delta m^2 $, $\delta A $) parameter space, for eq.(21).}
\end{figure}

\end{document}